\documentstyle[11pt]{article}
\begin{document}

\title{Velocity Curve Analysis of the Spectroscopic Binary Stars NSV
223, AB And, V2082 Cyg, HS Her, V918 Her, BV Dra, BW Dra, V2357 Oph,
and YZ Cas by the Non-linear Least Squares}
\author{K. Karami$^{1,2,3}$,\thanks{E-mail: karami@iasbs.ac.ir}\\
R. Mohebi${^1}$,\thanks{E-mail: rozitamohebi@yahoo.com}\\M. M. Soltanzadeh${^1}$,\thanks{E-mail: msoltanzadeh@uok.ac.ir}\\
$^{1}$\small{Department of Physics, University of Kurdistan,
Pasdaran St., Sanandaj, Iran}\\$^{2}$\small{Research Institute for
Astronomy
$\&$ Astrophysics of Maragha (RIAAM), Maragha, Iran}\\
$^{3}$\small{Institute for Advanced Studies in Basic Sciences
(IASBS), Gava Zang, Zanjan, Iran}}

\maketitle

\begin{abstract}
Using measured radial velocity data of nine double lined
spectroscopic binary systems NSV 223, AB And, V2082 Cyg, HS Her,
V918 Her, BV Dra, BW Dra, V2357 Oph, and YZ Cas, we find
corresponding orbital and spectroscopic elements via the method
introduced by Karami \& Mohebi (2007a) and Karami \& Teimoorinia
(2007). Our numerical results are in good agreement with those
obtained by others using more traditional methods.
\end{abstract} \noindent{Key words.~~~stars: binaries: eclipsing --- stars: binaries: spectroscopic}

\section{Introduction}
\label{intro} Determining the orbital elements of binary stars helps
us to obtain fundamental information, such as the masses and radii
of individual stars, that has an important role in understanding the
present state and evolution of many interesting stellar objects.
Analysis of both light and radial velocity (hereafter RV) curves,
derived from photometric and spectroscopic observations,
respectively, yields a complete set of basic absolute parameters.
One historically well-known method to analyze the RV curve is that
of Lehmann-Filh\'{e}s (cf. Smart, 1990).   In the present paper we
use the method introduced by Karami $\&$ Mohebi (2007a) (=KM2007a)
and Karami $\&$ Teimoorinia (2007) (=KT2007), to obtain orbital
parameters of the nine double-lined spectroscopic binary systems:
NSV 223, AB And, V2082 Cyg, HS Her, V918 Her, BV Dra, BW Dra, V2357
Oph, and YZ Cas.

The NSV 223 is a contact system of the A type and the mass ratio is
believed to be small (Rucinski et al. 2003a,b). The large
semiamplitudes ,$K_i$, suggest that the orbital inclination is close
to $90^\circ$. The spectral type is $F7V$ and the period is
$0.366128$ days (Rucinski et al. 2003a,b). The AB And is a contact
binary. The spectral type is $G8V$. The period is $0.3318919$ days.
It is suggested that observed period variability may be a result of
the orbital motion in a wide triple system. The third body should
then have to be a white dwarf (Pych et al. 2003,2004). V2082 Cyg is
most probably an A-type contact binary with a period of $0.714084$
days. The spectral type is $F2V$ (Pych et al. 2003,2004). In HS Her,
the effective temperatures were found to be $T_1=15200\pm750K$ and
$T_2=7600\pm400K$ for the primary and secondary stars, respectively
(Cakirli et al. 2007). The secondary component appears to rotate
more slowly. The presence of a third body physically bound to the
eclipsing pair has been suggested by many investigators. The two
component are located near the zero-age main sequence, with age of
about $32$ Myr (Cakirli et al. 2007). It is classified as an
Algol-type eclipsing binary and single-lined spectroscopic binary.
The spectral type of more massive primary component is $B4.5V$. The
effective temperature is about $15200\pm750K$ for the primary
component and $7600\pm400K$ for the secondary component. The period
is $1.6374316$ days (Cakirli et al. 2007). The V918 Her is an A-type
contact binary. The spectral classification is A7V. The period of
this binary star is $0.57481$ days (Pych et al. 2003,2004). The BV
Dra  and BW Dra have a circular orbit. From spectrophotometry the
components of BV Dar are classified as F9 and F8 while the
components of BW Dra are G3 and G0. The period of BV Dra is
$0.350066$ days and for BW Dra is $0.292166$ days (Batten \& Wenxian
1986). The V2357 Oph was classified as a pulsating star with a
period of $0.208$ days. The spectral type is G5V (Rucinski et al.
2003a,b). The spectral type of primary component of YZ Cas is A1V
and for the secondary is F7V. The period of this binary is
$4.4672235$ days (Lacy 1981).

This paper is organized as follows. In Sect. 2, we give a brief
review of the method of KM2007a and KT2007. In Sect. 3, the
numerical results are reported, while the conclusions are given in
Section 4.

\section{A brief review on the method of KM2007a and KT2007}
One may show that the radial acceleration scaled by the period is
obtained as
\begin{eqnarray}
P\ddot{Z}&=&\frac{-2\pi
K}{(1-e^2)^{3/2}}\sin\Big(\cos^{-1}(\frac{\dot{Z}}{K}-e\cos\omega)\Big)
\nonumber
\\&\times&\Big\{1+e\cos\Big(-\omega+\cos^{-1}(\frac{\dot{Z}}{K}-e\cos\omega)\Big)\Big\}^2,
\label{pz:}
\end{eqnarray}
where the dot denotes the time derivative, $e$ is the eccentricity,
$\omega$ is the longitude of periastron and $\dot{Z}$ is the radial
velocity of system with respect to the center of mass. Also
\begin{eqnarray}
K=\Large \frac{2\pi}{P}\frac{a\sin i}{\sqrt{1-e^2}},\label{K}
\end{eqnarray}
where $P$ is the period of motion, $a$ is the semimajor axis of the
orbit and inclination $i$ is the angle between the line of sight and
the normal of the orbital plane.

Equation (\ref{pz:}) describes a nonlinear relation,
$P\ddot{Z}=P\ddot{Z}(\dot{Z},K,e,\omega)$, in terms of the orbital
elements $K$, $e$ and $\omega$. Using the nonlinear regression of
Eq. (\ref{pz:}), one can estimate the parameters $K,~e$ and
$\omega$, simultaneously. Also one may show that the adopted
spectroscopic elements, i.e. $m_p/m_s$, $m_p\sin^3 i$ and $m_s\sin^3
i$, are related to the orbital parameters.
\section{Numerical Results}
\begin{figure}
\includegraphics{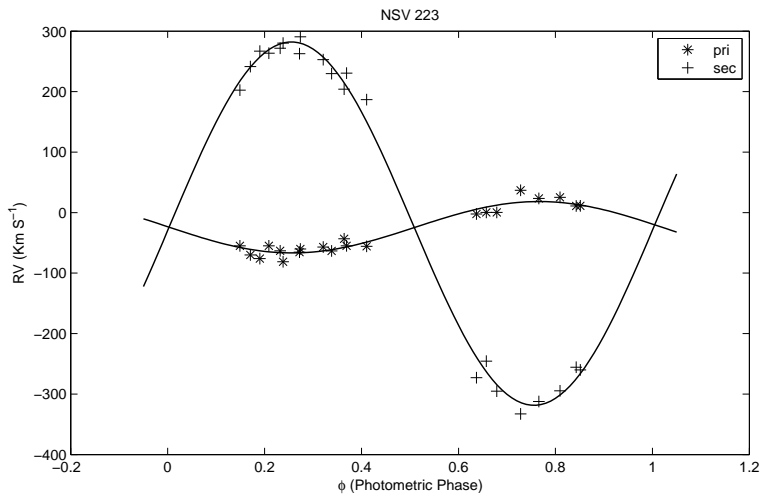}
      \vspace{9.8cm}
      \caption[]{Radial velocities of the primary and secondary components of
      NSV 223 plotted against the photometric phase. The observational data have been deduced from Rucinski et al. (2003a,b).}
         \label{NSV_RV}
   \end{figure}

\begin{figure}
\includegraphics{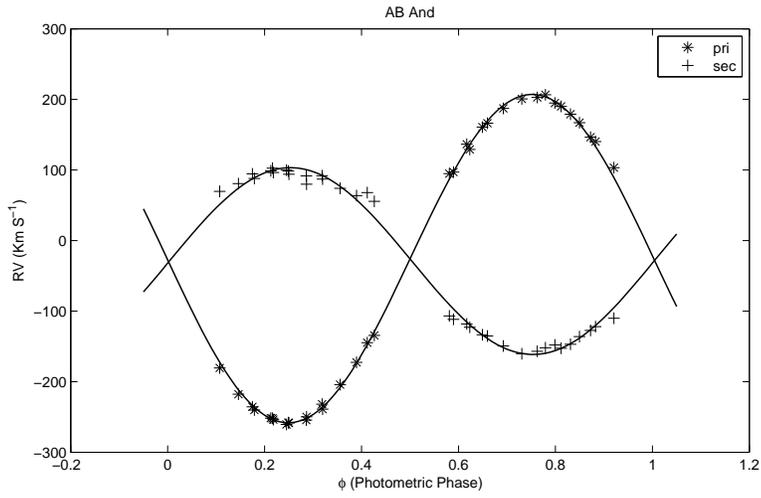}
      \vspace{9.8cm}
      \caption[]{Same as Fig. \ref{NSV_RV}, for AB And. The observational data have been measured by Pych et al. (2003,2004).}
         \label{AB_RV}
   \end{figure}

\begin{figure}
\includegraphics{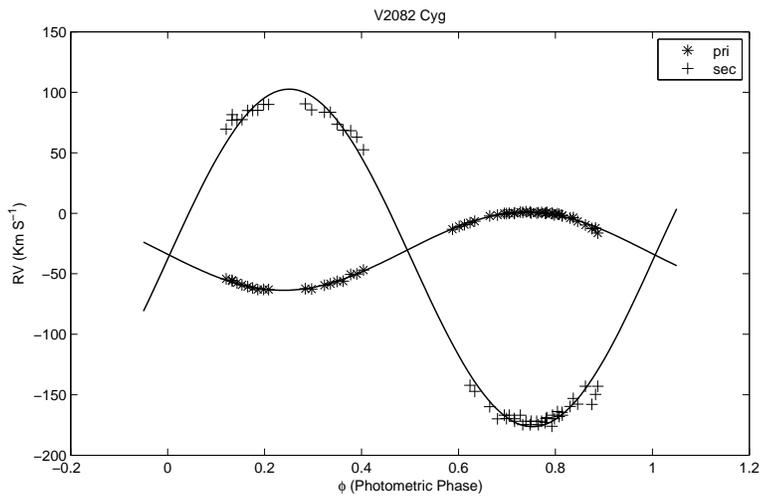}
      \vspace{9.8cm}
      \caption[]{Same as Fig. \ref{NSV_RV}, for V2082 Cyg.
      The observational data have been derived from Pych et al. (2003,2004).}
         \label{V2082_RV}
   \end{figure}

\begin{figure}
\includegraphics{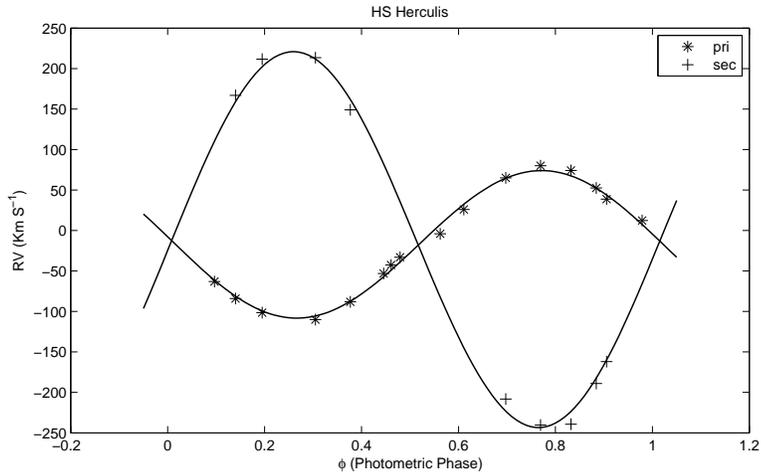}
      \vspace{9.8cm}
      \caption[]{Same as Fig. \ref{NSV_RV}, for HS Her. The observational data are from Cakirli et al. (2007).}
         \label{HS_RV}
   \end{figure}

\begin{figure}
\includegraphics{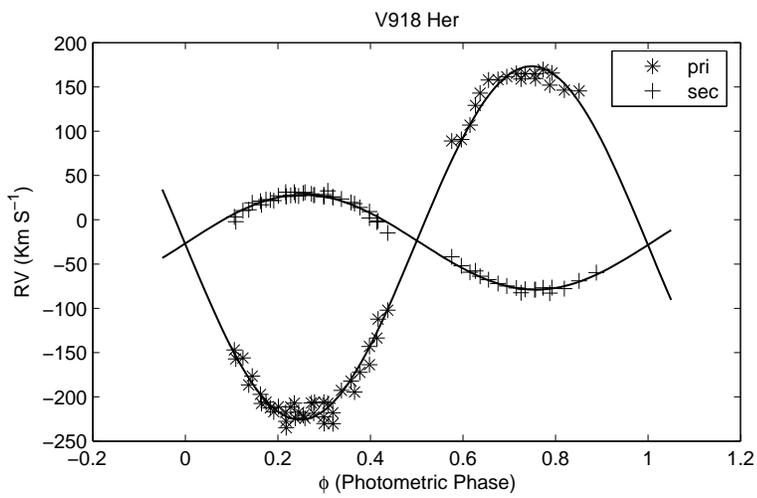}
      \vspace{7cm}
      \caption[]{Same as Fig. \ref{NSV_RV}, for V918 Her.
      The observational data have been derived from Pych et al.
      (2003,2004).}
         \label{V918_RV}
   \end{figure}

\begin{figure}
\includegraphics{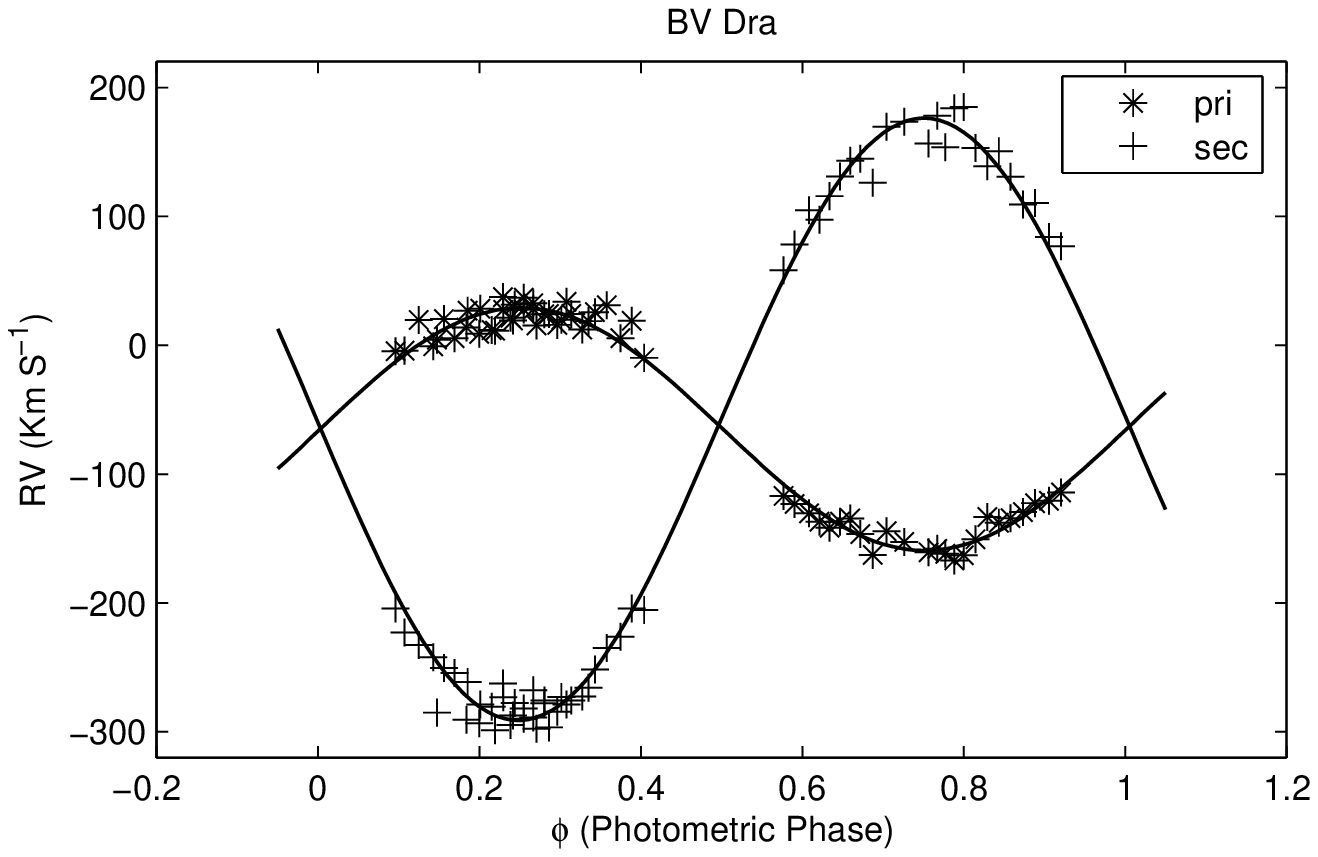}
      \vspace{7cm}
      \caption[]{Same as Fig. \ref{NSV_RV}, for BV Dra.
      The observational data have been derived from Batten \& Wenxian (1986).}
         \label{BV_RV}
   \end{figure}

\begin{figure}
\includegraphics{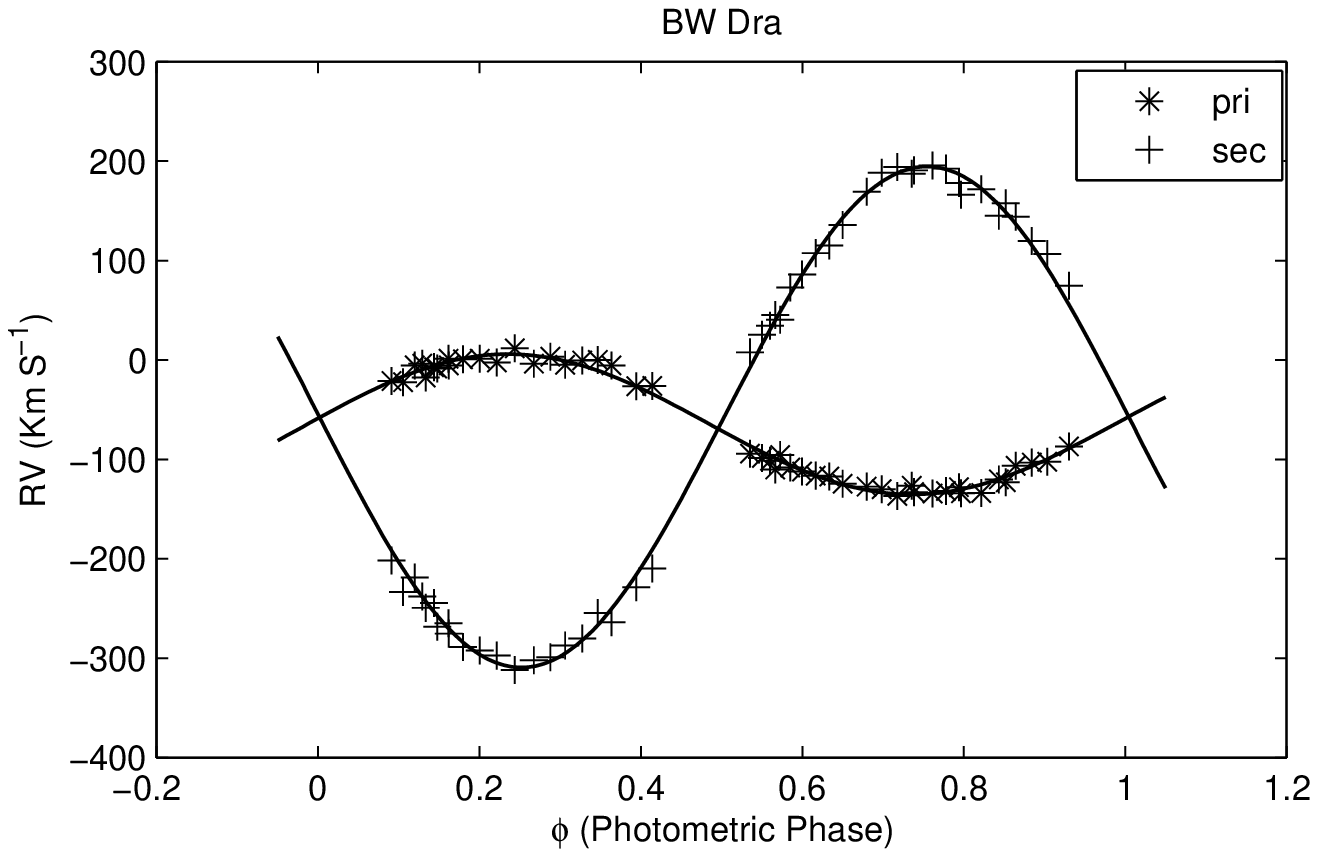}
      \vspace{7cm}
      \caption[]{Same as Fig. \ref{NSV_RV}, for BW Dra.
      The observational data have been derived from Batten \& Wenxian (1986).}
         \label{BW_RV}
   \end{figure}

\begin{figure}
\includegraphics{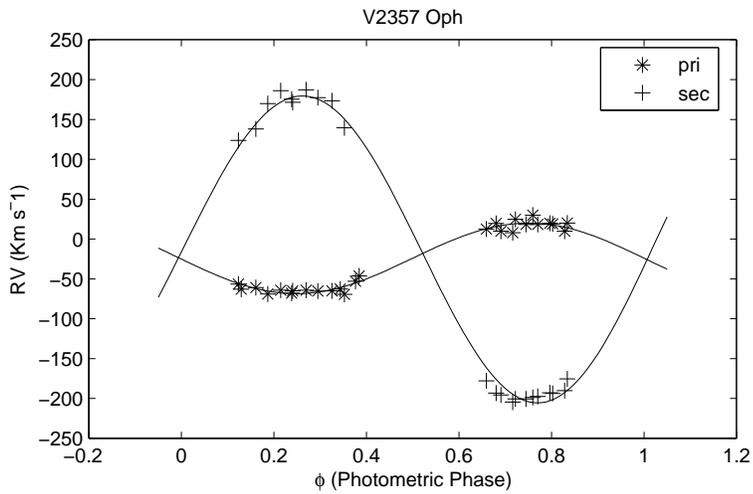}
      \vspace{7cm}
      \caption[]{Same as Fig. \ref{NSV_RV}, for V2357 Oph.
      The observational data have been derived from Rucinski et al.
      (2003a,b).}
         \label{V2357_RV}
   \end{figure}

\begin{figure}
\includegraphics{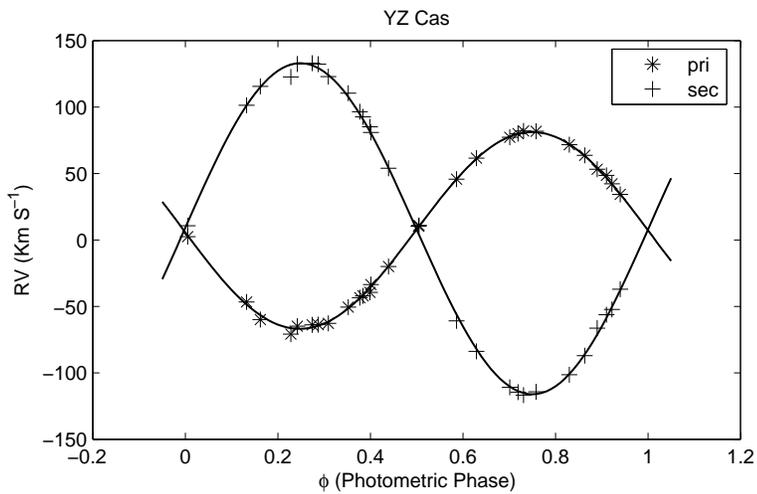}
      \vspace{7cm}
      \caption[]{Same as Fig. \ref{NSV_RV}, for YZ Cas.
      The observational data have been derived from Lacy (1981).}
         \label{YZ_RV}
   \end{figure}

Here we use the method of KM2007a and KT2007 to derive both the
orbital and combined elements for the nine different double lined
spectroscopic systems NSV 223, AB And, V2082 Cyg, HS Her, V918 Her,
BV Dra, BW Dra, V2357 Oph, and YZ Cas. Using measured radial
velocity data of the two components of these systems obtained by
Rucinski et al. (2003a,b) for NSV 223 and V2357 Oph, Pych et al.
(2003,2004) for AB And, V2082 Cyg, and V918 Her, Cakirli et al.
(2007) for HS Her, Batten \& Wenxian (1986) for BV Dra and BW Dra,
Lacy (1981) for YZ Cas, the fitted velocity curves are plotted in
terms of the photometric phase in Figs. \ref{NSV_RV} to \ref{YZ_RV}.

\begin{figure}
 \includegraphics{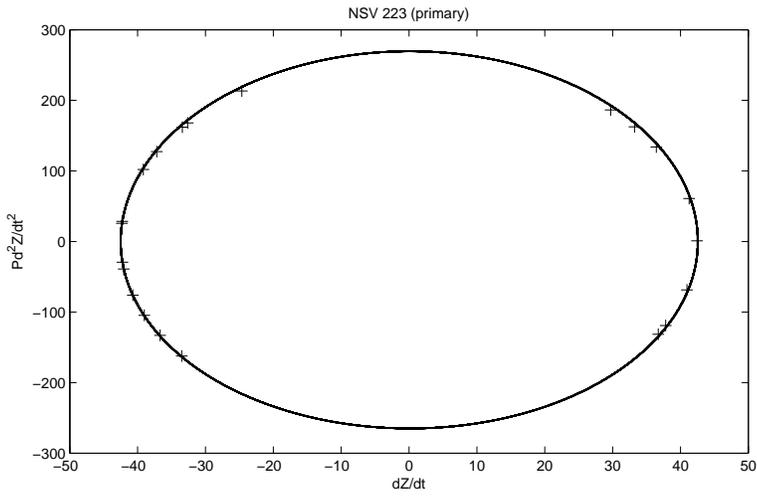}
      \vspace{9.8cm}
      \caption[]{The radial acceleration scaled by the period versus the radial velocity of the primary
      component of NSV 223. The solid curve is obtained from the non-linear regression of Eq. (\ref{pz:}).
      The plus points are
      the experimental data.}
         \label{NSV_pri}
   \end{figure}

\begin{figure}
\includegraphics{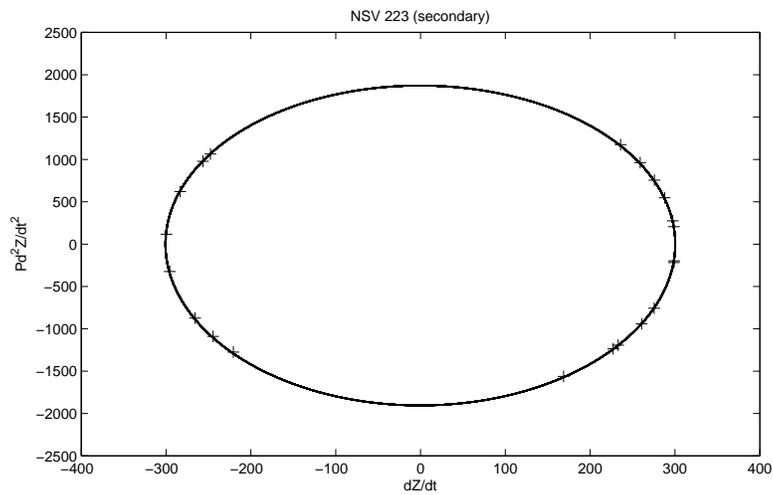}
      \vspace{9.8cm}
      \caption[]{Same as Fig. \ref{NSV_pri}, for the secondary component of NSV 223.}
         \label{NSV_sec}
   \end{figure}

\begin{figure}
 \includegraphics{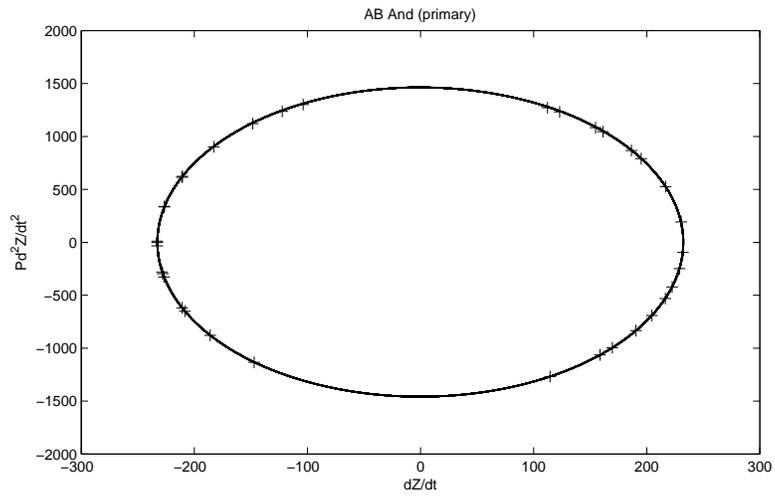}
      \vspace{9.8cm}
      \caption[]{Same as Fig. \ref{NSV_pri}, for the primary component of AB And.}
         \label{AB_pri}
   \end{figure}

\begin{figure}
\includegraphics{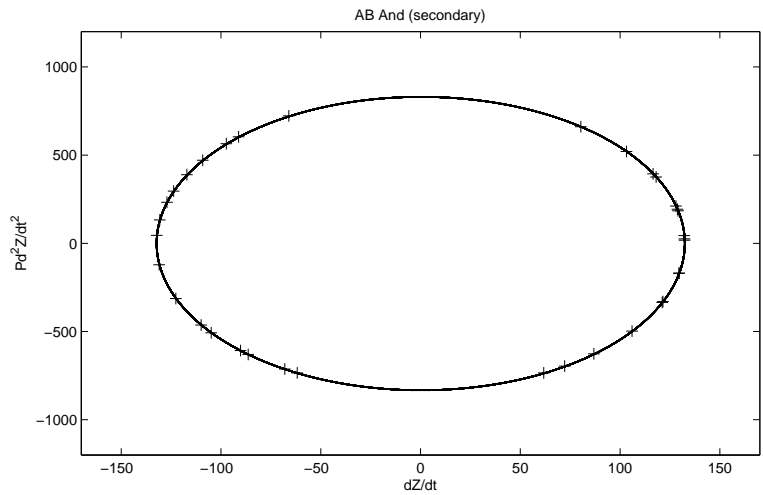}
      \vspace{9.8cm}
      \caption[]{Same as Fig. \ref{NSV_pri}, for the secondary component of AB And.}
         \label{AB_sec}
   \end{figure}

\begin{figure}
 \includegraphics{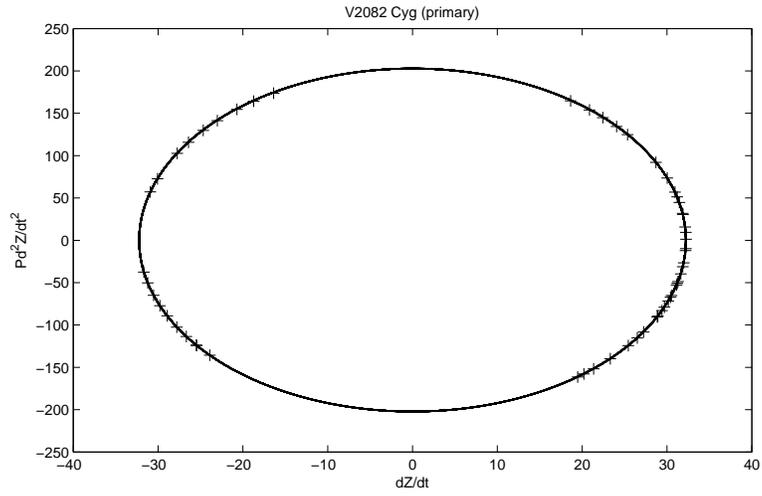}
      \vspace{9.8cm}
      \caption[]{Same as Fig. \ref{NSV_pri}, for the primary component of V2082 Cyg.}
         \label{V2082_pri}
   \end{figure}

\begin{figure}
\includegraphics{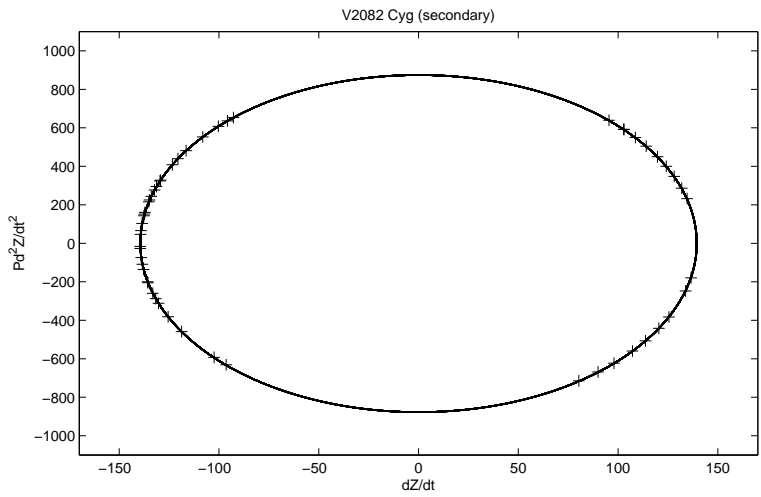}
      \vspace{9.8cm}
      \caption[]{Same as Fig. \ref{NSV_pri}, for the secondary component of V2082 Cyg.}
         \label{V2082_sec}
   \end{figure}

\begin{figure}
 \includegraphics{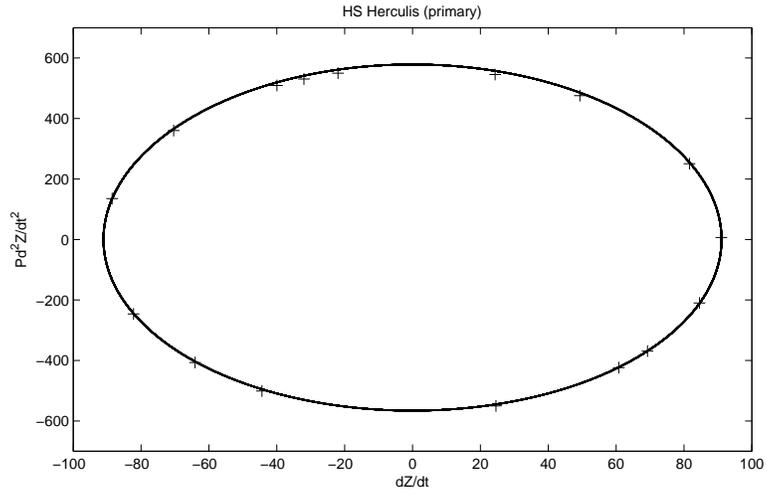}
      \vspace{9.8cm}
      \caption[]{Same as Fig. \ref{NSV_pri}, for the primary component of HS Her.}
         \label{HS_pri}
   \end{figure}

\begin{figure}
\includegraphics{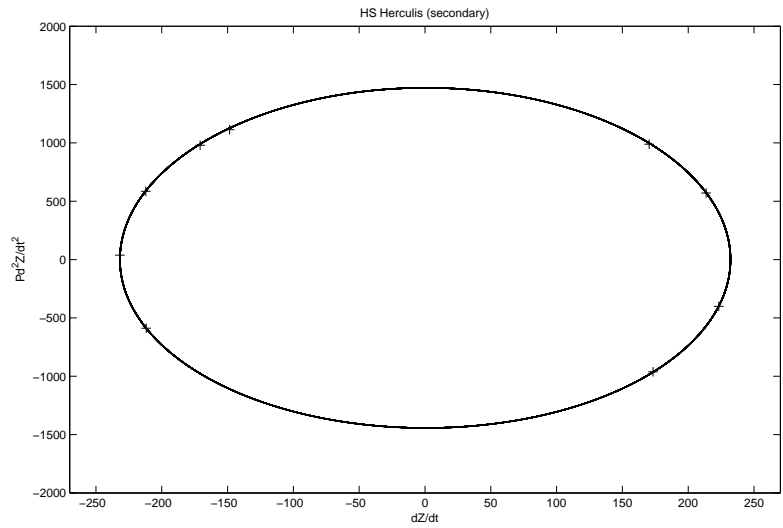}
      \vspace{11.5cm}
      \caption[]{Same as Fig. \ref{NSV_pri}, for the secondary component of HS Her.}
         \label{HS_sec}
   \end{figure}

\begin{figure}
\includegraphics{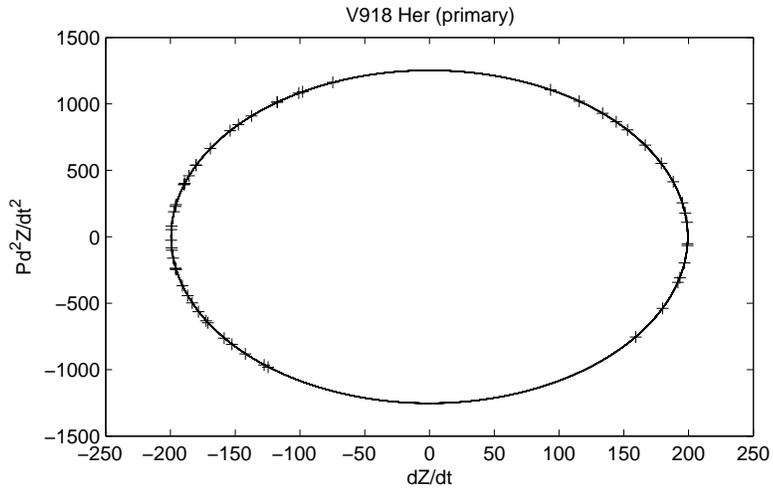}
      \vspace{7cm}
      \caption[]{Same as Fig. \ref{NSV_pri}, for the primary component of V918
      Her.}
         \label{V918_pri}
   \end{figure}

\begin{figure}
\includegraphics{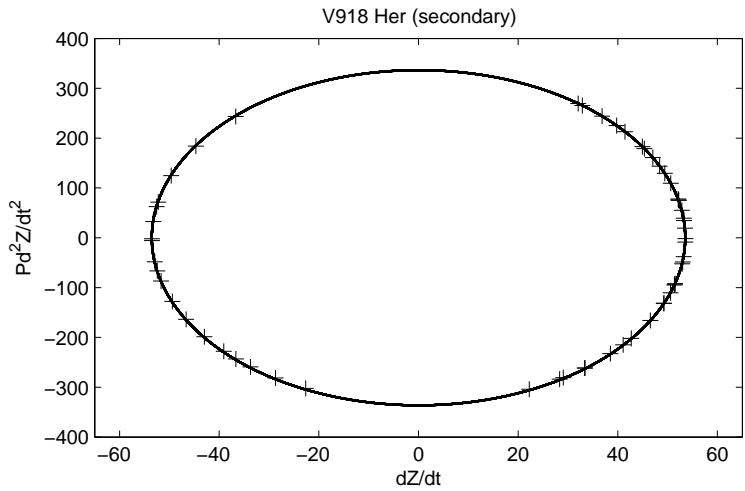}
      \vspace{7cm}
      \caption[]{Same as Fig. \ref{NSV_pri}, for the secondary component of V918 Her.}
         \label{V918_sec}
   \end{figure}

\begin{figure}
\includegraphics{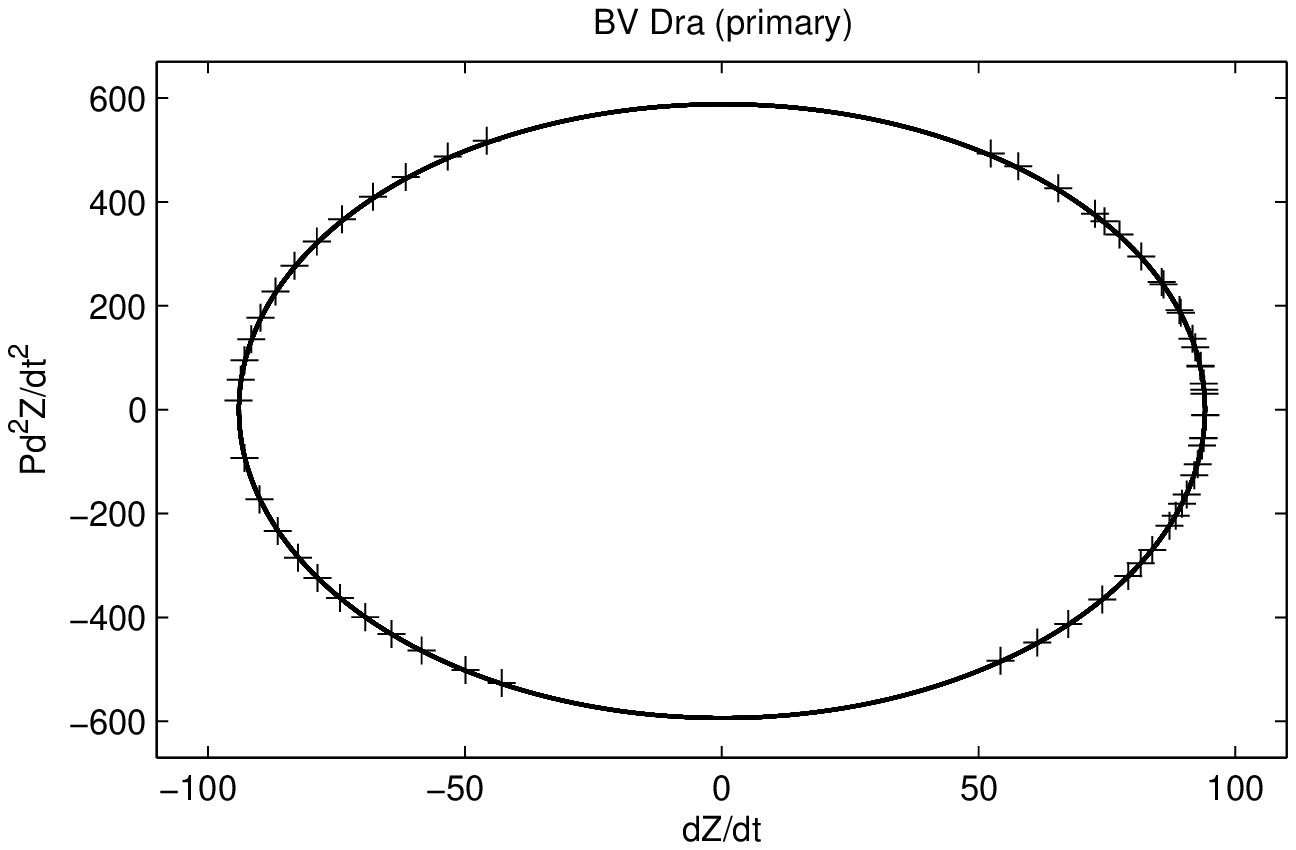}
      \vspace{7cm}
      \caption[]{Same as Fig. \ref{NSV_pri}, for the primary component of BV
      Dra.}
         \label{BV_pri}
   \end{figure}

\begin{figure}
\includegraphics{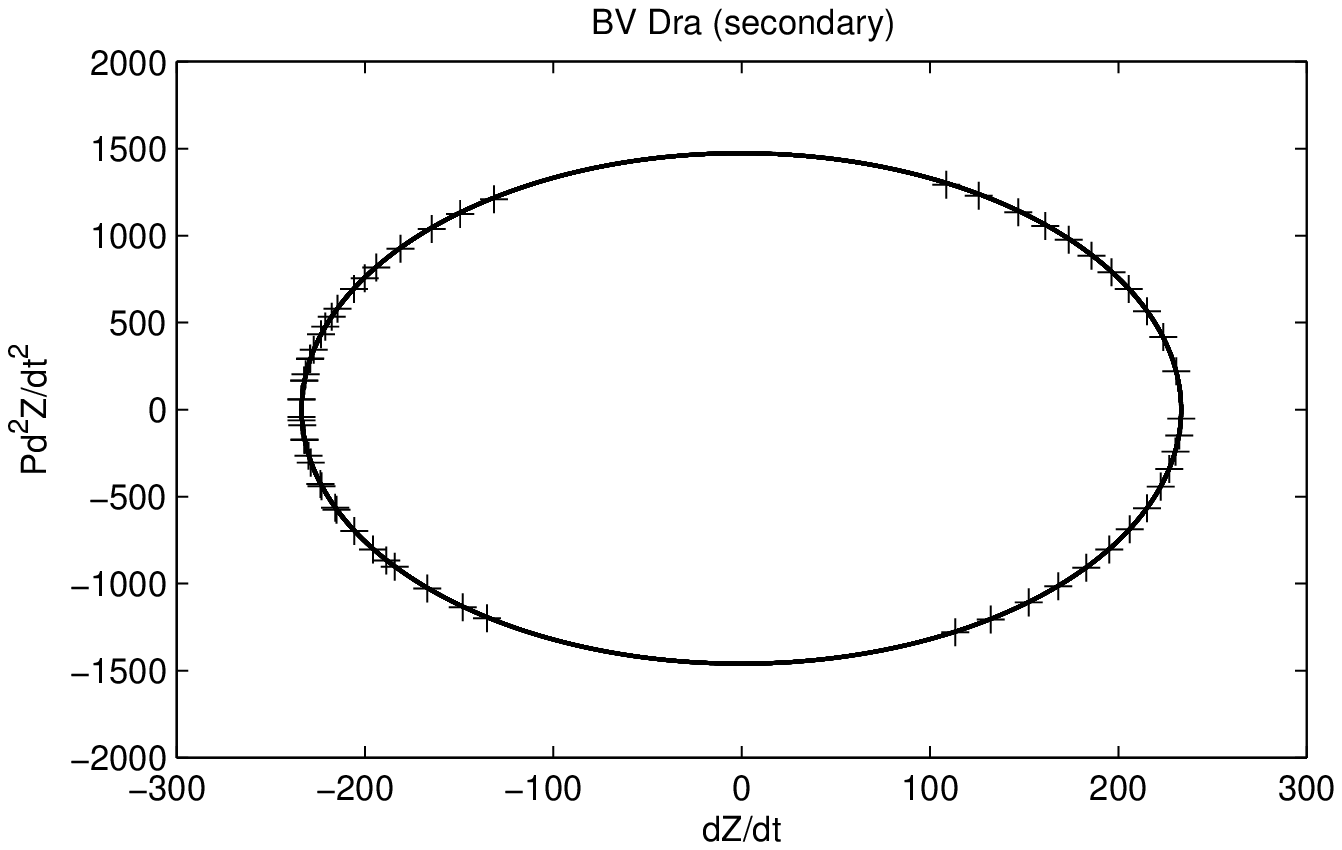}
      \vspace{7cm}
      \caption[]{Same as Fig. \ref{NSV_pri}, for the secondary component of BV
      Dra.}
         \label{BV_sec}
   \end{figure}

\begin{figure}
\includegraphics{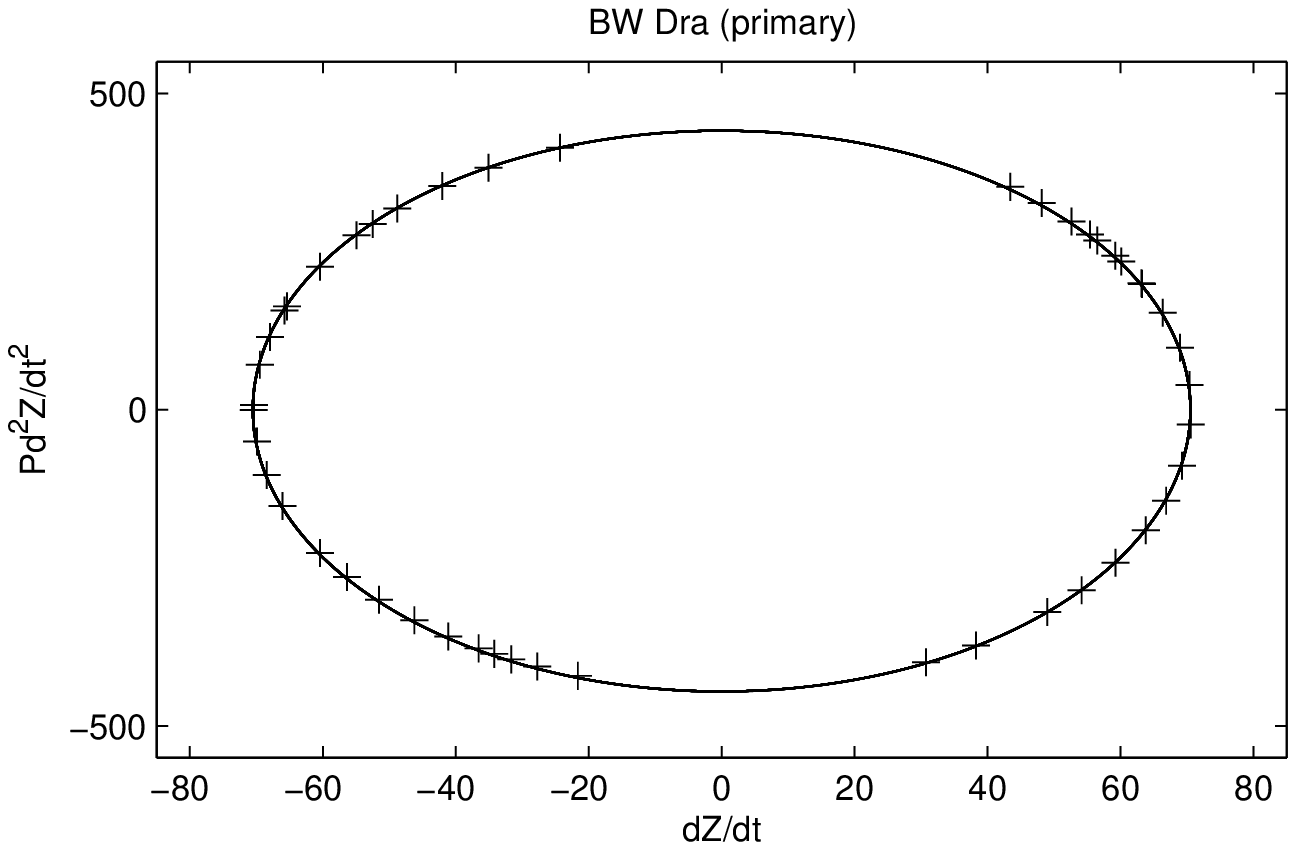}
      \vspace{7cm}
      \caption[]{Same as Fig. \ref{NSV_pri}, for the primary component of BW Dra.}
         \label{BW_pri}
   \end{figure}

\begin{figure}
\includegraphics{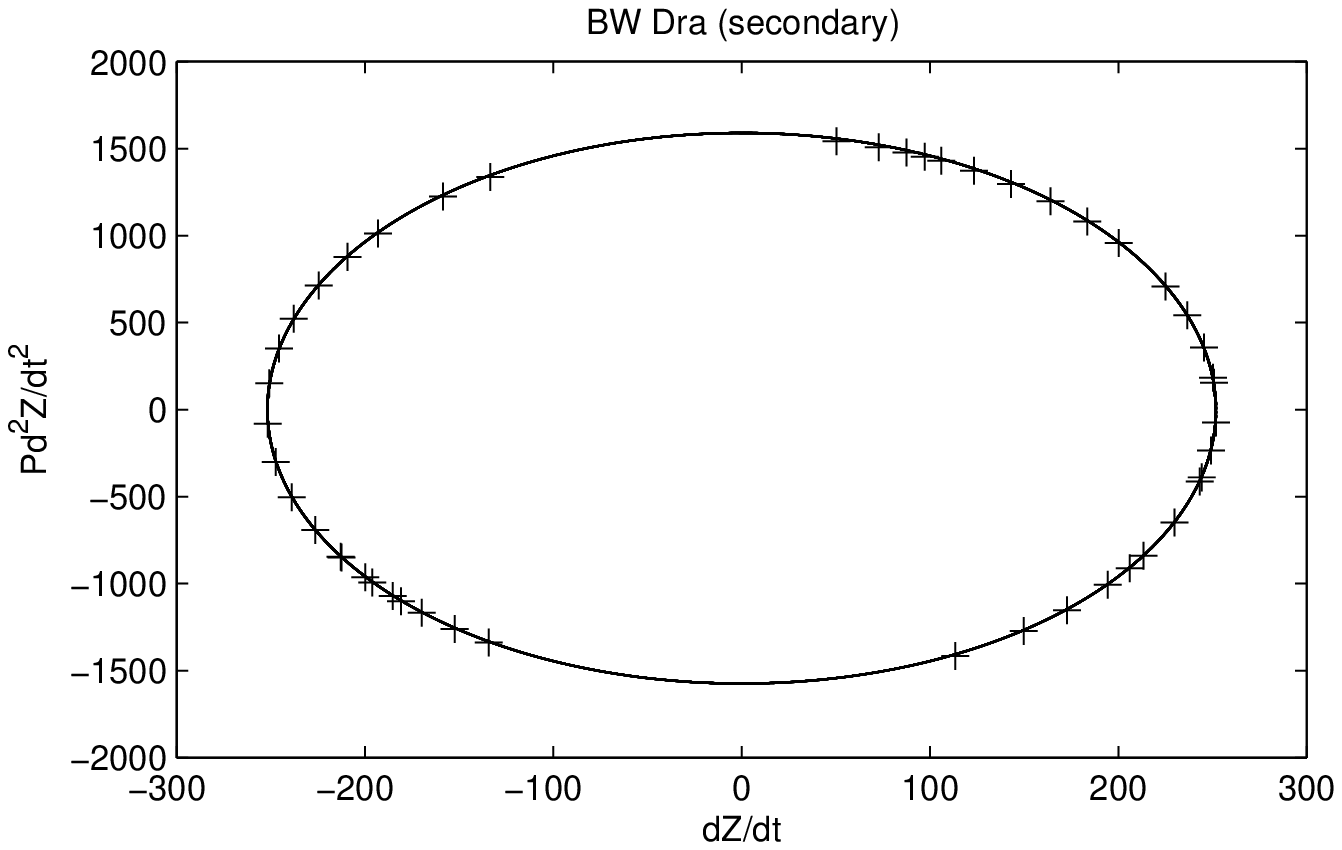}
      \vspace{7cm}
      \caption[]{Same as Fig. \ref{NSV_pri}, for the secondary component of BW
      Dra.}
         \label{BW_sec}
   \end{figure}

\begin{figure}
\includegraphics{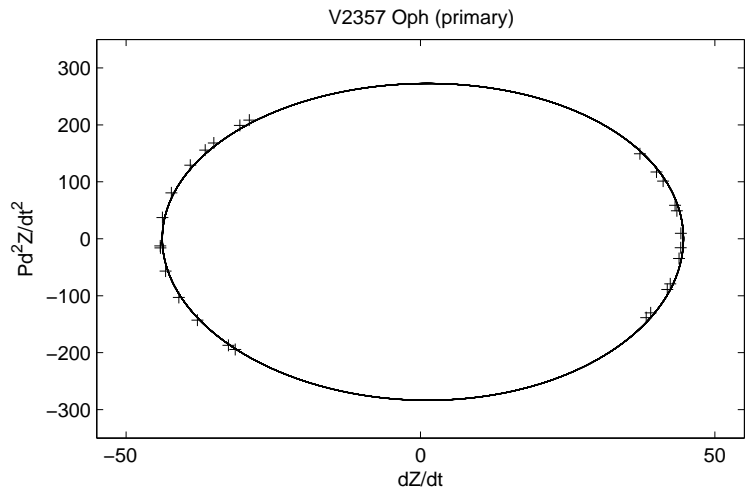}
      \vspace{7cm}
      \caption[]{Same as Fig. \ref{NSV_pri}, for the primary component of V2357
      Oph.}
         \label{V2357_pri}
   \end{figure}

\begin{figure}
\includegraphics{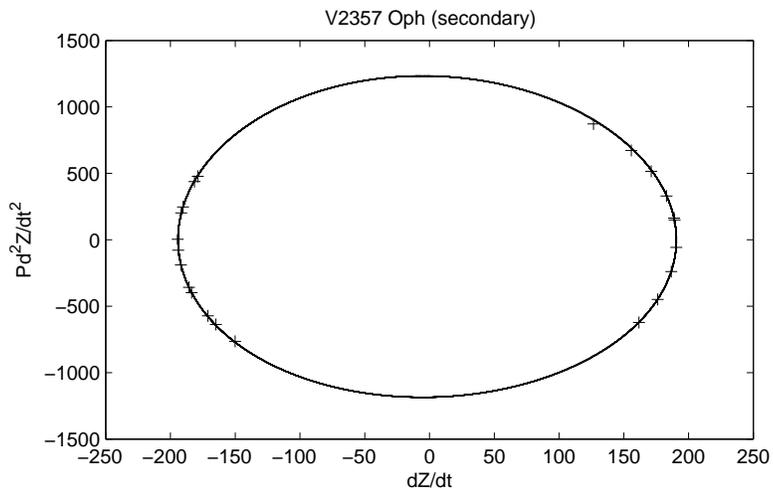}
      \vspace{7cm}
      \caption[]{Same as Fig. \ref{NSV_pri}, for the secondary component of V2357 Oph.}
         \label{V2357_sec}
   \end{figure}

\begin{figure}
\includegraphics{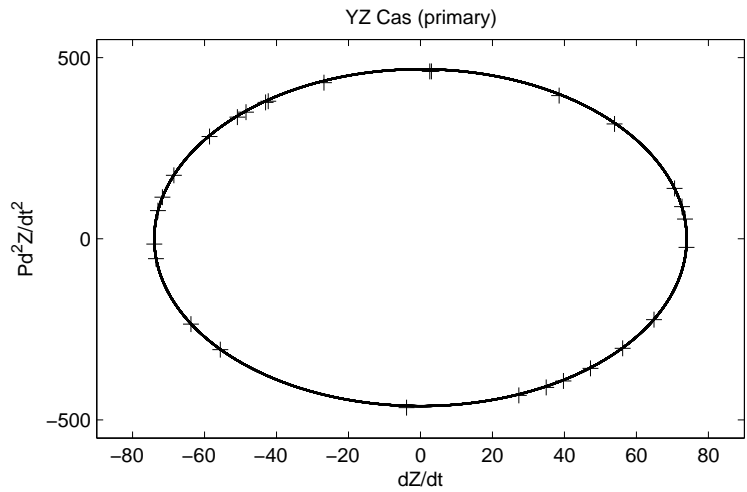}
      \vspace{7cm}
      \caption[]{Same as Fig. \ref{NSV_pri}, for the primary component of YZ Cas.}
         \label{YZ_pri}
   \end{figure}

\begin{figure}
\includegraphics{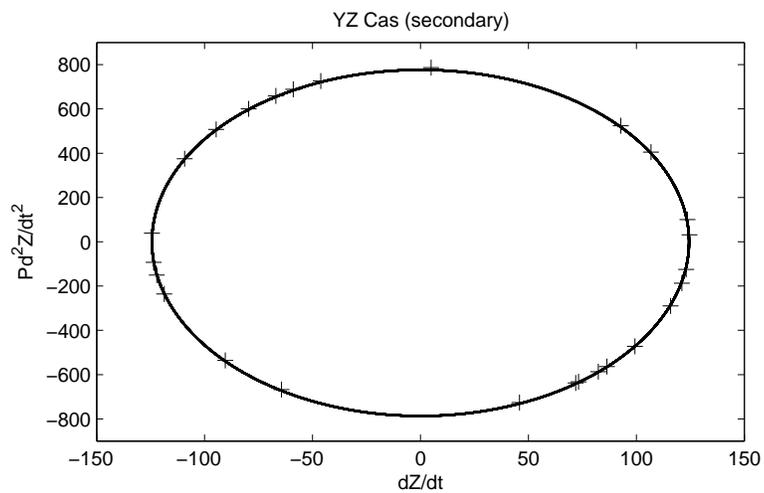}
      \vspace{7cm}
      \caption[]{Same as Fig. \ref{NSV_pri}, for the secondary component of YZ
      Cas.}
         \label{YZ_sec}
   \end{figure}

Figures \ref{NSV_pri} to \ref{YZ_sec} show the radial acceleration
scaled by the period versus the radial velocity for the primary and
secondary components of NSV 223, AB And, V2082 Cyg, HS Her, V918
Her, BV Dra, BW Dra, V2357 Oph, and YZ Cas, respectively. The solid
closed curves are results obtained from the non-linear regression of
Eq. (\ref{pz:}), which their good coincidence with the measured data
yields to derive the optimized parameters $K$, $e$ and $\omega$. The
apparent closeness of these curves to ellipses is due to the low, or
zero, eccentricities of the corresponding orbits. For a well-defined
eccentricity, the acceleration-velocity curve shows a noticeable
deviation from a regular ellipse (see Karami $\&$ Mohebi 2007b).

The orbital parameters, $K$, $e$ and $\omega$, obtaining from the
non-linear least squares of Eq. (\ref{pz:}) and the combined
spectroscopic elements including $m_p\sin^3i$, $m_s\sin^3i$,
$(a_p+a_s)\sin i$ and $m_p/m_s$ obtaining from the estimated
parameters $K$, $e$ and $\omega$ for the nine systems are tabulated
in Tables \ref{NSV_Orbit} to \ref{YZ_Combined}, respectively. The
velocity of the center of mass, $V_{{\rm cm}}$, is obtained by
calculating the areas above and below of the radial velocity curve.
Where these areas become equal to each other, the velocity of center
of mass is obtained. Tables \ref{NSV_Orbit} to \ref{YZ_Combined}
show that the results are in good accordance with the those obtained
by Rucinski et al. (2003a,b) for NSV 223 and V2357 Oph, Pych et al.
(2003,2004) for AB And, V2082 Cyg, and V918 Her, Cakirli et al.
(2007) for HS Her, Batten \& Wenxian (1986) for BV Dra and BW Dra,
and Lacy (1981) for YZ Cas.
\section{Conclusions}
Using the method introduced by KM2007a and KT2007, we obtain both
the orbital elements and the combined spectroscopic parameters of
nine double lined spectroscopic binary systems. Our results are in
good agreement with the those obtained by others using more
traditional methods. In a subsequent paper we intend to study the
other different systems.
\\
\\
\noindent{{\bf Acknowledgements}}. This work has been supported
financially by Research Institute for Astronomy $\&$ Astrophysics of
Maragha (RIAAM), Maragha, Iran.

\clearpage
\clearpage
\begin{table}
\begin{center}
\centering\caption[]{Orbital parameters of NSV 223.}
\begin{tabular}{lcc}\hline\noalign{\smallskip}
 &This Paper  & Rucinski et al. (2003a,b) \\\hline\noalign{\smallskip}
{\bf Primary} & \\
$V_{cm}\left(kms^{-1}\right)$ & $-21.31\pm0.98$ & $-21.66(2.15)$   \\
 $K_p\left(kms^{-1}\right)$&  $42.56\pm 0.03$ &$40.39(2.64)$   \\
 $e$& $0.004\pm0.001$ &--- \\
 $\omega(^\circ)$&$279.40\pm12.04$&--- \\
{\bf Secondary} &   \\
$V_{cm}\left(kms^{-1}\right)$ & $-21.31\pm0.98$ & $-21.66(2.15)$  \\
 $K_s\left(kms^{-1}\right)$&  $300.76\pm0.04$ & $297.98(4.18)$  \\
 $e$& $e_s=e_p$ & --- \\
$\omega(^\circ)$& $\omega_s=\omega_p-180^\circ$ &---
\\\hline\noalign{\smallskip}
\end{tabular}\\
\label{NSV_Orbit}
\end{center}
\end{table}

\begin{table}
\centering\caption[]{Spectroscopic elements of NSV 223.}
\begin{tabular}{lcc}\hline\noalign{\smallskip}
 Parameter  & This Paper & Rucinski et al. (2003a,b)  \\\hline\noalign{\smallskip}
 $m_p\sin^3i/M_\odot$ & $1.3447\pm0.0007$ & --- \\
$m_s\sin^3i/M_\odot$ & $0.1903\pm0.0002$ & --- \\
$\left(a_p+a_s\right)\sin i/R_\odot$ & $2.4834\pm0.0005$ & --- \\
$m_s/m_p$ & $0.1415\pm0.0001$ & $0.136(10)$ \\
$\left(m_p+m_s\right)\sin^3i/M_\odot$ & $1.535\pm0.001$& $1.47(9)$
\\\hline\noalign{\smallskip}
\end{tabular}\\
\label{NSV_Combined}
\end{table}
\begin{table}
\centering \caption[]{Same as Table \ref{NSV_Orbit}, for AB And}
\begin{tabular}{lccc}\hline\noalign{\smallskip}
 &This Paper  & Pych et al. (2003,2004)&\\
\hline\noalign{\smallskip}
 {\bf Primary} &   \\
$V_{cm} \left(kms^{-1}\right)$ & $-27.26\pm 0.66$ & $-27.53(0.67)$&  \\
 $K_p\left(kms^{-1}\right)$&  $232.69\pm0.02$ & $232.88(0.83)$ &  \\
 $e$& $0.00109\pm 0.00005$ &---&  \\
$\omega(^\circ)$& $230.69\pm 3.22$ &---& \\
{\bf Secondary} & \\
$V_{cm} \left(kms^{-1}\right)$ & $-27.26\pm0.66$ & $-27.53(0.67)$ &\\
 $K_s\left(kms^{-1}\right)$&  $132.43\pm0.01$ &$130.32(1.17)$ &\\
 $e$& $e_s=e_p$ &---& \\
 $\omega(^\circ)$&$\omega_s=\omega_p-180^\circ$&---&
\\\hline\noalign{\smallskip}
\end{tabular}\\
\label{AB_Orbit}
\end{table}

\begin{table}
\centering\caption[]{Same as Table \ref{NSV_Combined}, for AB And}
\begin{tabular}{lcc}\hline\noalign{\smallskip}
 Parameter  & This Paper & Pych et al. (2003,2004) \\\hline\noalign{\smallskip}
 $m_p\sin^3i/M_\odot$ & $0.6071\pm 0.0001$ & ---\\
$m_s\sin^3i/M_\odot$ & $1.0668\pm 0.0002$ & ---\\
$\left(a_p+a_s\right)\sin i/R_\odot$ & $2.3943\pm 0.0002$ & ---\\
$m_p/m_s$ & $0.5691\pm 0.0001$ & $0.560(7)$\\
$\left(m_p+m_s\right)\sin^3i/M_\odot$
&$1.6739\pm0.0004$&$1.648(20)$\\\hline\noalign{\smallskip}
\end{tabular}\\
\label{AB_Combined}
\end{table}
\clearpage
\begin{table}
\centering\caption[]{Same as Table \ref{NSV_Orbit}, for V2082 Cyg}
\begin{tabular}{lcc}\hline\noalign{\smallskip}
 &This Paper  & Pych et al. (2003,2004) \\\hline\noalign{\smallskip}
{\bf Primary} & \\
$V_{cm}\left(kms^{-1}\right)$ & $-34.19\pm0.39$ & $-34.12(0.58)$   \\
 $K_p\left(kms^{-1}\right)$&  $32.254\pm 0.003$ &$33.16(0.51)$   \\
 $e$& $0.0008\pm0.0005$ &--- \\
 $\omega(^\circ)$&$266.48\pm0.06$&--- \\
{\bf Secondary} &   \\
$V_{cm}\left(kms^{-1}\right)$ & $-34.19\pm0.39$ & $-34.12(0.58)$  \\
 $K_s\left(kms^{-1}\right)$&  $139.50\pm0.01$ & $139.38(0.99)$  \\
 $e$& $e_s=e_p$ & --- \\
$\omega(^\circ)$& $\omega_s=\omega_p-180^\circ$ &---
\\\hline\noalign{\smallskip}
\end{tabular}\\
\label{V2082_Orbit}
\end{table}

\begin{table}
\centering\caption[]{Same as Table \ref{NSV_Combined}, for V2082
Cyg}
\begin{tabular}{lcc}\hline\noalign{\smallskip}
 Parameter  & This Paper & Pych et al. (2003,2004)  \\\hline\noalign{\smallskip}
 $m_p\sin^3i/M_\odot$ & $0.3045\pm0.0001$ & --- \\
$m_s\sin^3i/M_\odot$ & $0.07039\pm0.00002$ & --- \\
$\left(a_p+a_s\right)\sin i/R_\odot$ & $2.4232\pm0.0002$ & --- \\
$m_s/m_p$ & $0.23121\pm0.00004$ & $0.238(5)$ \\
$\left(m_p+m_s\right)\sin^3i/M_\odot$ & $0.3749\pm0.0001$&
$0.380(7)$
\\\hline\noalign{\smallskip}
\end{tabular}\\
\label{V2082_Combined}
\end{table}
\begin{table}
\centering\caption[]{Same as Table \ref{NSV_Orbit}, for HS Her}
\begin{tabular}{lcc}\hline\noalign{\smallskip}
  &This paper  & Cakirli et al. (2007)\\\hline\noalign{\smallskip}
{\bf Primary} &   \\
$V_{cm}\left(kms^{-1}\right)$ & $-14.24\pm0.87$ & $-12.8\pm1.9$\\
 $K_p\left(kms^{-1}\right)$&  $91.15\pm0.02$ & $93.4\pm2.8$\\
 $e$& $0.048\pm0.001$ & $0.05\pm0.01$\\
$\omega(^\circ)$& $268.66\pm 0.03$ & ---\\
{\bf Secondary} & \\
$V_{cm}\left(kms^{-1}\right)$ & $-14.24\pm0.87$ & $-12.8\pm1.9$\\
 $K_s\left(kms^{-1}\right)$&  $232.13\pm0.01$ &$239.4\pm3.2$\\
 $e$& $e_s=e_p$ &$0.05\pm0.01$\\
 $\omega(^\circ)$&$\omega_s=\omega_p-180^\circ$ &\\\hline\noalign{\smallskip}
\end{tabular}\\
\label{HS_Orbit}
\end{table}

\begin{table}
\centering\caption[]{Same as Table \ref{NSV_Combined}, for HS Her}
\begin{tabular}{lccc}\hline\noalign{\smallskip}
  Parameter  & This study & Cakirli et al. (2007)\\\hline\noalign{\smallskip}
 $m_p\sin^3i/M_\odot$ & $4.116\pm0.001$ & ---\\
$m_s\sin^3i/M_\odot$ & $1.616\pm0.001$ &---\\
$\left(a_p+a_s\right)\sin i/R_\odot$ & $10.458\pm0.001$ &$10.76\pm0.50$\\
$m_s/m_p$ & $0.3927\pm0.0001$ & $0.39\pm0.05$
\\\hline\noalign{\smallskip}
\end{tabular}\\                                                                                                                                                                                                                                                                                                                                                                                                                    
\label{HS_Combined}
\end{table}
\clearpage
\begin{table}
\centering \caption[]{Same as Table \ref{NSV_Orbit}, for V918 Her}
\begin{tabular}{lccc}\hline\noalign{\smallskip}
&This Paper  & Pych et al. (2003,2004)&\\
\hline\noalign{\smallskip}
{\bf Primary} &   \\
$V_{cm} \left(kms^{-1}\right)$ & $-25.76\pm 0.78$ & $-25.72(0.74)$&  \\
$K_p\left(kms^{-1}\right)$&  $199.54\pm0.01$ & $199.37(1.72)$ &  \\
$e$& $0.0002\pm 0.0001$ &---&  \\
$\omega(^\circ)$& $315.14\pm 12.17$ &---& \\
{\bf Secondary} & \\
$V_{cm} \left(kms^{-1}\right)$ & $-25.76\pm0.78$ & $-25.72(0.74)$ &\\
$K_s\left(kms^{-1}\right)$&  $53.577\pm0.001$ &$53.93(0.49)$ &\\
$e$& $e_s=e_p$ &---& \\
$\omega(^\circ)$&$\omega_s=\omega_p-180^\circ$&---&
\\\hline\noalign{\smallskip}
\end{tabular}\\
\label{V918_Orbit}
\end{table}
\begin{table}
\centering\caption[]{Same as Table \ref{NSV_Combined}, for V918 Her}
\begin{tabular}{lcc}\hline\noalign{\smallskip}
Parameter  & This Paper & Pych et al. (2003,2004)
\\\hline\noalign{\smallskip}
$m_p\sin^3i/M_\odot$ & $0.20442\pm 0.00002$ & ---\\
$m_s\sin^3i/M_\odot$ & $0.7613\pm 0.0001$ & ---\\
$m_p/m_s$ & $0.26851\pm 0.00001$ & $0.271(5)$\\
$\left(m_p+m_s\right)\sin^3i/M_\odot$
&$0.9657\pm0.0001$&$0.968(21)$\\\hline\noalign{\smallskip}
\end{tabular}\\
\label{V918_Combined}
\end{table}
\begin{table}
\centering\caption[]{Same as Table \ref{NSV_Orbit}, for BV Dra}
\begin{tabular}{lcc}\hline\noalign{\smallskip}
 &This Paper  & Batten \& Wenxian (1986) \\\hline\noalign{\smallskip}
{\bf Primary} & \\
$V_{cm}\left(kms^{-1}\right)$ & $-61.23\pm0.77$ & $-65.2\pm1.2$   \\
 $K_p\left(kms^{-1}\right)$&  $94.119\pm 0.002$ &$93.9\pm1.4$   \\
 $e$& $0.0025\pm0.0003$ &--- \\
 $\omega(^\circ)$&$256.06\pm11.18$&--- \\
{\bf Secondary} &   \\
$V_{cm}\left(kms^{-1}\right)$ & $-61.23\pm0.77$ & $-57.7\pm1.7$  \\
 $K_s\left(kms^{-1}\right)$&  $139.50\pm0.01$ & $139.38(0.99)$  \\
 $e$& $e_s=e_p$ & --- \\
$\omega(^\circ)$& $\omega_s=\omega_p-180^\circ$ &---
\\\hline\noalign{\smallskip}
\end{tabular}\\
\label{BV_Orbit}
\end{table}
\begin{table}
\centering\caption[]{Same as Table \ref{NSV_Combined}, for BV Dra}
\begin{tabular}{lcc}\hline\noalign{\smallskip}
 Parameter  & This Paper & Batten \& Wenxian (1986)  \\\hline\noalign{\smallskip}
$a_pSini/10^6$& $0.453\pm0.001$& $0.452\pm0.007$\\
$a_sSini/10^6$& $1.1248\pm0.0001$& $1.124\pm0.010$\\
$m_p\sin^3i/M_\odot$ & $0.9106\pm0.0002$ &$0.911\pm.020$\\
$m_s\sin^3i/M_\odot$ & $0.36679\pm0.00006$ &$0.366\pm0.010$ \\
$m_s/m_p$ & $0.40278\pm0.00004$ & $0.402\pm0.007$
\\\hline\noalign{\smallskip}
\end{tabular}\\
\label{BV_Combined}
\end{table}
\clearpage
\begin{table}
\centering \caption[]{Same as Table \ref{NSV_Orbit}, for BW Dra}
\begin{tabular}{lccc}\hline\noalign{\smallskip}
&This Paper  & Batten \& Wenxian (1986)&\\
\hline\noalign{\smallskip}
{\bf Primary} &   \\
$V_{cm} \left(kms^{-1}\right)$ & $-61.14\pm 0.73$ & $-64.2\pm0.9$&  \\
$K_p\left(kms^{-1}\right)$&  $70.59\pm0.03$ & $70.5\pm1.1$ &  \\
$e$& $e_p=e_s$ &---  \\
$\omega(^\circ)$& $270.58\pm0.44$ &--- \\
{\bf Secondary} & \\
$V_{cm} \left(kms^{-1}\right)$ & $-61.14\pm0.73$ & $-57.4\pm1.4$ &\\
$K_s\left(kms^{-1}\right)$&  $252.018\pm0.005$ &$251.5\pm1.7$ &\\
$e$& $0.0026\pm 0.0001$ &--- \\
$\omega(^\circ)$&$\omega_s=\omega_p-180^\circ$&---&
\\\hline\noalign{\smallskip}
\end{tabular}\\
\label{BW_Orbit}
\end{table}
\begin{table}
\centering\caption[]{Same as Table \ref{NSV_Combined}, for BW Dra}
\begin{tabular}{lcc}\hline\noalign{\smallskip}
Parameter  & This Paper & Batten \& Wenxian (1986)
\\\hline\noalign{\smallskip}
$a_p\sin i/10^6$& $0.2836\pm0.0001$& $0.283\pm0.004$\\
$a_s\sin i/10^6$& $1.01249\pm0.00002$& $1.010\pm0.007$\\
$m_p\sin^3i/M_\odot$ & $0.7939\pm 0.0001$ & $0.791\pm0.015$\\
$m_s\sin^3i/M_\odot$ & $0.2224\pm0.0001$ & $0.222\pm0.005$\\
$m_p/m_s$ &
$0.2801\pm0.0001$&$0.280\pm0.005$
\\\hline\noalign{\smallskip}
\end{tabular}\\
\label{BW_Combined}
\end{table}
\begin{table}
\begin{center}
\centering\caption[]{Orbital parameters of V2357 Oph.}
\begin{tabular}{lcc}\hline\noalign{\smallskip}
 &This Paper  & Rucinski et al. (2003a,b) \\\hline\noalign{\smallskip}
{\bf Primary} & \\
$V_{cm}\left(kms^{-1}\right)$ & $-17.79\pm0.63$ & $-19.12(1.64)$   \\
 $K_p\left(kms^{-1}\right)$&  $44.29\pm 0.07$ &$44.12(1.63)$   \\
 $e$& $e_p=e_s$ &--- \\
 $\omega(^\circ)$&$227.12\pm 0.23$&--- \\
{\bf Secondary} &   \\
$V_{cm}\left(kms^{-1}\right)$ & $-17.79\pm0.63$ & $-19.12(1.64)$  \\
 $K_s\left(kms^{-1}\right)$&  $192.52\pm0.02$ & $190.93(2.90)$  \\
 $e$& $0.0134\pm0.0004$ & --- \\
$\omega(^\circ)$& $\omega_s=\omega_p-180^\circ$ &---
\\\hline\noalign{\smallskip}
\end{tabular}\\
\label{V2357_Orbit}
\end{center}
\end{table}
\begin{table}
\centering\caption[]{Spectroscopic elements of V2357 Oph.}
\begin{tabular}{lcc}\hline\noalign{\smallskip}
 Parameter  & This Paper & Rucinski et al. (2003a,b)  \\\hline\noalign{\smallskip}
 $m_p\sin^3i/M_\odot$ & $0.4647\pm0.0004$ & --- \\
$m_s\sin^3i/M_\odot$ & $0.1069\pm0.0002$ & --- \\
$m_s/m_p$ & $0.2300\pm0.0004$ & $0.231(10)$ \\
$\left(m_p+m_s\right)\sin^3i/M_\odot$ & $0.5716\pm0.0007$&
$0.560(32)$
\\\hline\noalign{\smallskip}
\end{tabular}\\
\label{V2357_Combined}
\end{table}
\begin{table}
\centering \caption[]{Same as Table \ref{NSV_Orbit}, for YZ Cas}
\begin{tabular}{lccc}\hline\noalign{\smallskip}
&This Paper  &Lacy (1981)&\\
\hline\noalign{\smallskip}
{\bf Primary} &   \\
$V_{cm} \left(kms^{-1}\right)$ & $7.78\pm 0.58$ & $8.14\pm0.06$&  \\
$K_p\left(kms^{-1}\right)$&  $74.039\pm0.002$ & $73.35\pm0.21$ &  \\
$e$& $0.0034\pm 0.0001$ &$0.0\pm0.003$&  \\
$\omega(^\circ)$& $270.078\pm 0.395$ &---& \\
{\bf Secondary} & \\
$V_{cm} \left(kms^{-1}\right)$ & $7.78\pm0.58$ & $8.14\pm0.06$ &\\
$K_s\left(kms^{-1}\right)$&  $124.49\pm0.05$ &$125.7\pm0.5$ &\\
$e$& $e_s=e_p$ &---& \\
$\omega(^\circ)$&$\omega_s=\omega_p-180^\circ$&---&
\\\hline\noalign{\smallskip}
\end{tabular}\\
\label{YZ_Orbit}
\end{table}
\begin{table}
\centering\caption[]{Same as Table \ref{NSV_Combined}, for YZ Cas}
\begin{tabular}{lcc}\hline\noalign{\smallskip}
Parameter  & This Paper & Lacy (1981) \\\hline\noalign{\smallskip}
$a_p\sin i/10^6$& $4.548\pm0.003$& $4.508\pm0.012$\\
$a_s\sin i/10^6$& $7.6470\pm0.0001$& $7.727\pm0.031$\\
$m_p/M_\odot$ & $2.274\pm0.002$ & $2.31\pm0.01$\\
$m_s/M_\odot$ & $1.352\pm0.001$ & $1.35\pm0.02$
\\\hline\noalign{\smallskip}
\end{tabular}\\
\label{YZ_Combined}
\end{table}

\end{document}